\documentclass[journal]{IEEEtranTIE}
\usepackage{graphicx}
\usepackage{cite}
\usepackage{picinpar}
\usepackage{amsmath}
\usepackage{url}
\usepackage{flushend}
\usepackage[latin1]{inputenc}
\usepackage{colortbl}
\usepackage{soul}
\usepackage{multirow}
\usepackage{pifont}
\usepackage{color}
\usepackage{alltt}
\usepackage{enumerate}
\usepackage{siunitx}
\usepackage{epstopdf}
\usepackage{pbox}
\usepackage{threeparttable}
\usepackage{algorithm2e}

\usepackage{xcolor}

\begin{document}
\title{Missing-Class-Robust  Domain Adaptation by Unilateral Alignment for Fault Diagnosis}
\author{
	\vskip 1em
	{
	Qin Wang,
	Gabriel Michau,
	Olga Fink, \emph{Member, IEEE}
	}

	\thanks{
		
		{
		All authors are with the chair of Intelligent Maintenance Systems, IBI, ETH Zurich, Switzerland (e-mail: \{qwang, gmichau, ofink\}@ethz.ch). 
		
		This work was supported by the Swiss National Science Foundation (SNSF) Grant no. PP00P2\_176878.
		
		}
	}
}

\maketitle
	
\begin{abstract}
Domain adaptation aims at improving model performance by leveraging the learned knowledge in the source domain and transferring it to the target domain. Recently, domain adversarial methods have been particularly successful in alleviating the distribution shift between the source and the target domains. However, these methods assume an identical label space between the two domains. This assumption imposes a significant limitation for real applications since the target training set may not contain the complete set of classes.  
We demonstrate in this paper that the performance of domain adversarial methods can be vulnerable to an incomplete target label space during training. To overcome this issue, we propose a two-stage unilateral alignment approach. The proposed methodology makes use of the inter-class relationships of the source domain and aligns unilaterally the target to the source domain. The benefits of the proposed methodology are first evaluated on the MNIST$\rightarrow$MNIST-M adaptation task. The proposed methodology is also evaluated on a fault diagnosis task, where the problem of missing fault types in the target training dataset is common in practice.  Both experiments demonstrate the effectiveness of the proposed methodology.

\end{abstract}


\markboth{preprint}%
{}

\definecolor{limegreen}{rgb}{0.2, 0.8, 0.2}
\definecolor{forestgreen}{rgb}{0.13, 0.55, 0.13}
\definecolor{greenhtml}{rgb}{0.0, 0.5, 0.0}

\section{Introduction}

\IEEEPARstart{I}n recent years, deep learning methods~\cite{aizenberg2001multi, lecun2015deep, goodfellow2016deep} have achieved some remarkable results on various tasks~\cite{he2016identity,devlin2019bert, girshick2015fast}. However, the methods require not only large training datasets, but also labels to learn the relevant patterns in the data. This data-intensive nature of deep learning methods and particularly the requirement of labels, which can be expensive or even impossible to acquire, has limited their utilization in practical applications. In addition, the trained models  usually don't generalize well if a distribution shift is encountered between training and test data.  

Unsupervised domain adaptation techniques~\cite{fernando2013unsupervised,ganin2014unsupervised,long2015learning, sun2015subspace } provide a promising solution to alleviate both challenges: missing labels and domain shift. Domain adaptation aims at leveraging unlabeled target data to improve the model's generalization ability in the target domain. It allows knowledge transfer from a source domain to a different but related target domain~\cite{pan2011domain}. Recently, adversarial domain adaptation approaches~\cite{ganin2014unsupervised, tzeng2015simultaneous, ganin2016domain, tzeng2017adversarial, luo2017label, long2018conditional} have significantly improved domain adaption performance by aligning source and target data in an adversarial way and enforcing domain-invariant features in the latent space. 

\begin{figure}
    	\centering
	{\includegraphics[width=0.9\columnwidth]{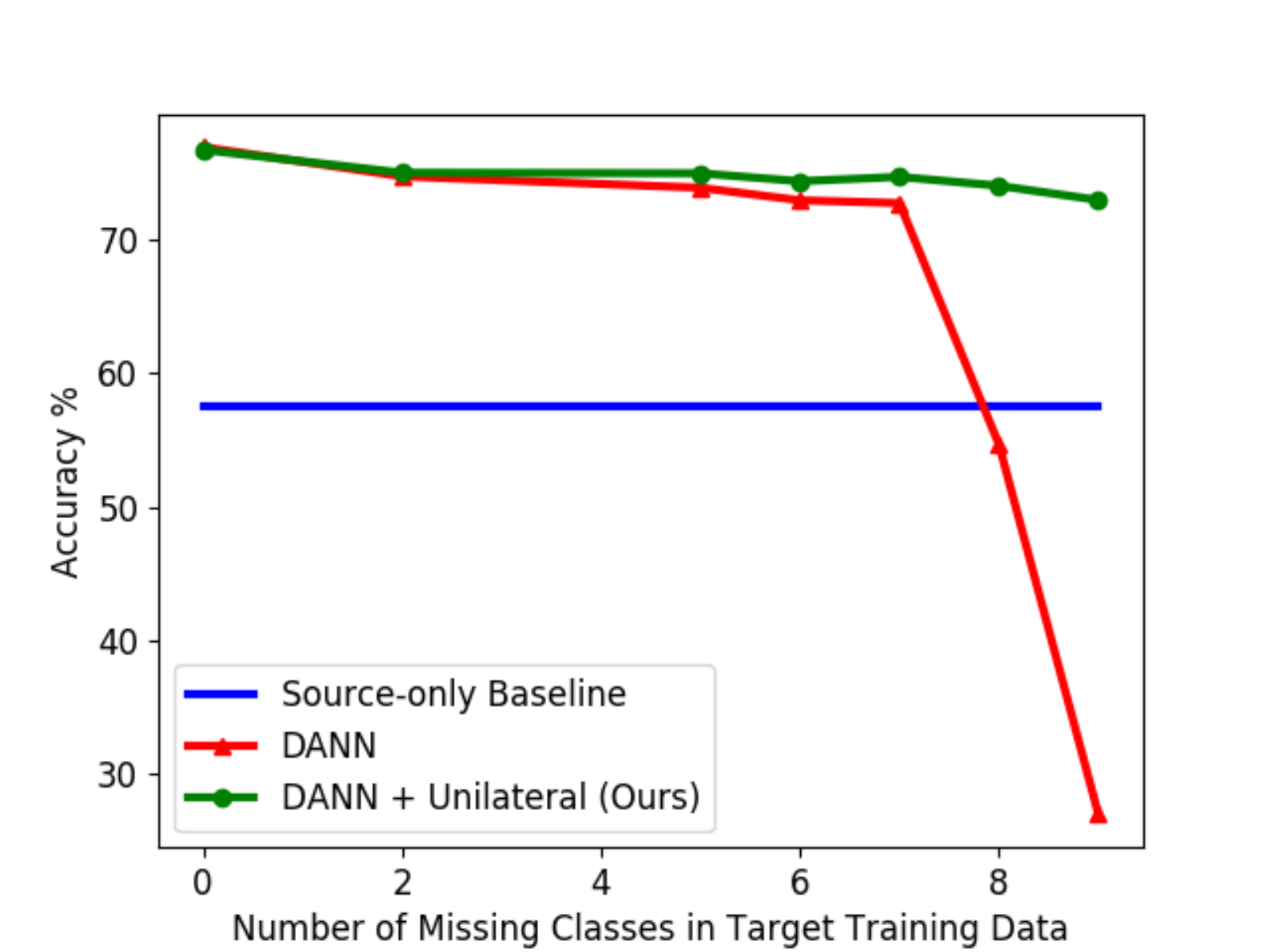}}
	
	\caption{Effect of different numbers of missing classes in the target training set on adversarial domain adaptation approach on the MNIST$\rightarrow$ MNIST-M  task:  Baseline model trained using only source data (Blue).  Model trained using traditional adversarial domain adaptation (Red). The proposed unilateral adaptation that mitigates the misalignment risk caused by the missing classes (Green) .}
	\label{miss}
\end{figure}
These unsupervised domain adaptation methods relax the need of labels on the target domain by transferring knowledge across domains. However, they typically assume that the label space of source and target data is identical~\cite{saito2018open}. This assumption imposes a significant limitation for real applications since the training dataset may not contain the complete set of classes. 
For example, in the industrial fault diagnosis, it would be beneficial to be able to transfer the source knowledge without having to wait for the occurrence of every possible fault in the target domain.  :

Direct adversarial domain alignment results in a large uncertainty on the model performance in case of missing classes in the target domain. The domain alignment is, in this case, performed between a source domain with a complete set of labels and a target training set containing only a subset of labels. The alignment results in a negative effect on the model performance on the missing classes in the target training set. One additional drawback of the direct domain alignment is the negative effect on the inter-class relationship. When domain adaptation techniques are applied on the non-identical label space, the alignment effect is different among the classes that are present in the target domain during training and the missing classes. This means that the inter-class relationship for the aligned domains is likely  to be distorted in an unpredictable way from the original inter-class relationship in the source domain. It is expected that, this misalignment effect is larger when there are more missing classes in the target domain. We demonstrate an example of this phenomenon using the MNIST$\rightarrow$ MNIST-M adaptation task in Figure~\ref{miss} with a varying number of missing classes.

In order to mitigate this negative alignment effect and better transfer the learned inter-class relationships  from source  to target domain, we propose to unilaterally align target domain towards the source domain instead of aligning both to an unknown intermediate space. The ultimate goal of the proposed approach is to make adversarial domain adaptation robust against missing classes in the target domain.

The intuition of the approach is to  make use of the discriminative information learned from source domain in a better way. Since we don't have access to samples of some classes in the target domain, we can only infer them by using the inter-class relationships we learned from the source domain. We argue that this approach will provide a more robust representation for the target domain.

The proposed method is a two-stage process: 1) We first train an anchor model based on  source data only and extract pre-trained source features. 2) By minimizing the distance between the source features and the pre-trained features, while performing source and target feature alignment, the target distribution is unilaterally transformed to match the source distribution. 

 We show the improvement on the MNIST$\rightarrow$ MNIST-M adaptation task in Figure~\ref{miss}  achieved by the proposed method with a varying number of missing classes in the target dataset.

To summarize, we propose a solution for the domain adaptation problem with missing classes in the target training data, while the model performance is still evaluated on all classes in the target domain. Similar to previous methods, we align the features in an adversarial way. However, different from previous methods, we align them unilaterally towards the source.   

The main contribution of this paper is the proposed methodology to preserve the inter-class relationships of the source domain \textbf{ by aligning the target domain distribution unilaterally towards the source domain} instead of jointly transforming the two domains to a common distorted latent space.  The proposed  methodology is particularly beneficial in problem setups with an incomplete set of class labels in the target training dataset.  

We validate the effectiveness of the proposed methods first on the MNIST$\rightarrow$MNIST-M transfer task.  
For this alignment task, the proposed methodology demonstrates its robustness to the incompleteness of the target label space and achieves a similar level of performance in case of missing classes as with the complete set of classes.  

To demonstrate the applicability of the proposed methodology in practical applications, we additionally evaluate our method on the task of transferring the learned knowledge between two different operating conditions for the fault diagnosis tasks on a bearing dataset.  By applying the proposed unilateral alignment methodology, we are able to improve diagnosis performance on the bearing dataset. 

\section{Related Work}
\subsection{Domain Adaptation}
Domain adaptation has been successfully applied in fields such as computer vision and natural language understanding~\cite{patel2015visual,csurka2017domain,pan2011domain, long2015learning, li2016revisiting, ganin2014unsupervised,  saito2018maximum}.  One common idea underlying different domain adaptation methods is the alleviation of the distribution discrepancy between source and target data, or in other words, they aim at aligning the source and target distributions. Different approaches have been proposed to address this task. \cite{pan2011domain} proposed to use transfer components analysis across domains. \cite{long2015learning} proposed to align the source and target distributions by minimizing the Maximum Mean Discrepancy (MMD). Driven by a similar motivation,  Adaptive Batch Normalization (AdaBN)~\cite{li2016revisiting} and AutoDial~\cite{carlucci2017autodial} align the distributions via modified batch normalization layers. Domain Adversarial Neural Networks (DANN)~\cite{ganin2014unsupervised} aim at aligning the distributions by using a domain discriminator and train the model adversarially in order to make the feature space indistinguishable for the different domains.

\subsection{Missing Class in Partial Domain Adaptation}
Another related topic is partial domain adaptation~\cite{cao2018partialcvpr}. In this setup, algorithms aim at solving the problem, where the target classes are a subset of source classes. \cite{zhang2018deep} proposed to use importance weighted adversarial networks to focus on shared classes. \cite{cao2018partial} alleviates negative transfer by down-weighing the data of outlier source classes. \cite{cao2019learning} proposed to learn domain-invariant representations across domains and a progressive weighting scheme.

In partial domain adaptation setup, the missing target classes are not evaluated during test time. This is the key difference with the setup used in our paper. 

\subsection{Domain Adaptation in Fault Diagnosis}
Missing class problem is especially severe when adapting domains or operating conditions for fault diagnosis problems.
Without considering missing classes, domain adaptation methods~\cite{zhang2017new, zhang2018adversarial, wang2019domain, li2019multi , michau2019, yang2019intelligent, 8643085} have recently been introduced to the fault diagnosis problems. Several approaches~\cite{lu2016deep, li2018cross} have been proposed to deal with missing-classes in the the context of fault diagnosis. However, they assume that the target training dataset contains exactly one class (the healthy condition). The main difference with our paper is that our proposed method is able to deal with different number of missing classes, meaning that the proposed method is more general as we don't assume the target training data all from healthy condition.

\section{Problem Description}
\label{sec:pd}
Formally, we mainly consider the following unsupervised domain adaptation task with missing classes.

\begin{itemize}
    \item Training data from source domain with all classes $$\mathcal{D}_s=\{(x_{s}^1, y_{s}^1), ...,(x_{s}^n, y_{s}^n)\},  y_{s}^i \in Y.$$
    \item Unlabeled training data from target domain with missing classes  $$\mathcal{D}_{t}=\{(x_{t}^1, y_{t}^1), ...,(x_{t}^m, y_{t}^m)\},  y_{t}^i  \in Y_{sub}, Y_{sub}\subset Y.$$%
    \item Test data from target domain with all classes $$\mathcal{D}_{test}=\{(x_{test}^1, y_{test}^1), ...,(x_{test}^k, y_{test}^k)\},  y_{test}^i \in Y,$$
\end{itemize}
where $Y$ is the complete set of classes and the labels $Y_{sub}$ of target training set only contains a subset of it. Note that for the target training set, it is unknown which classes belong to $Y_{sub}$. The only assumption is $Y_{sub}\subset Y$. Therefore, these samples are also part of the test set since their correct classification also needs to be evaluated. This evaluation of the classification accuracy of samples that were used for the alignment is in fact similar to that used by~\cite{li2018cross} and is common in transductive~\cite{gammerman1998learning, arnold2007comparative} domain adaptation problems.

\section{Proposed Methodology}
The missing classes in the target training set makes the direct application of standard domain adaptation methods difficult. The shared idea behind most methods is to transform both source and target into a shared feature space. Such an alignment requires sufficient support, from both source and target, and also from all classes. The lack of information on the target domain can lead to an unexpected behavior of the alignment. One of the potential issues is that the alignment is explicitly changing the distribution of the given classes in the target domain, while no supervised domain guidance is given for the missing classes in the target domain. This unbalanced alignment behavior between present and missing classes may distort the well-learnt source inter-class relationships, and thus make the alignment sub-optimal or even deteriorate the model performance. 

In order to fully leverage the limited healthy data from target domain and improve model performance on all classes, we propose a two-stage framework. We first learn a classification model in the source domain, and extract relevant features for source data. In the second stage, we apply domain adversarial adaptation techniques and align the  source and target data. We strengthen the alignment by making sure the alignment is unilateral, that is, forcing the aligned features to be as close as possible to those learned in the first step.

\begin{figure}
    	\centering
	{\includegraphics[width=\columnwidth]{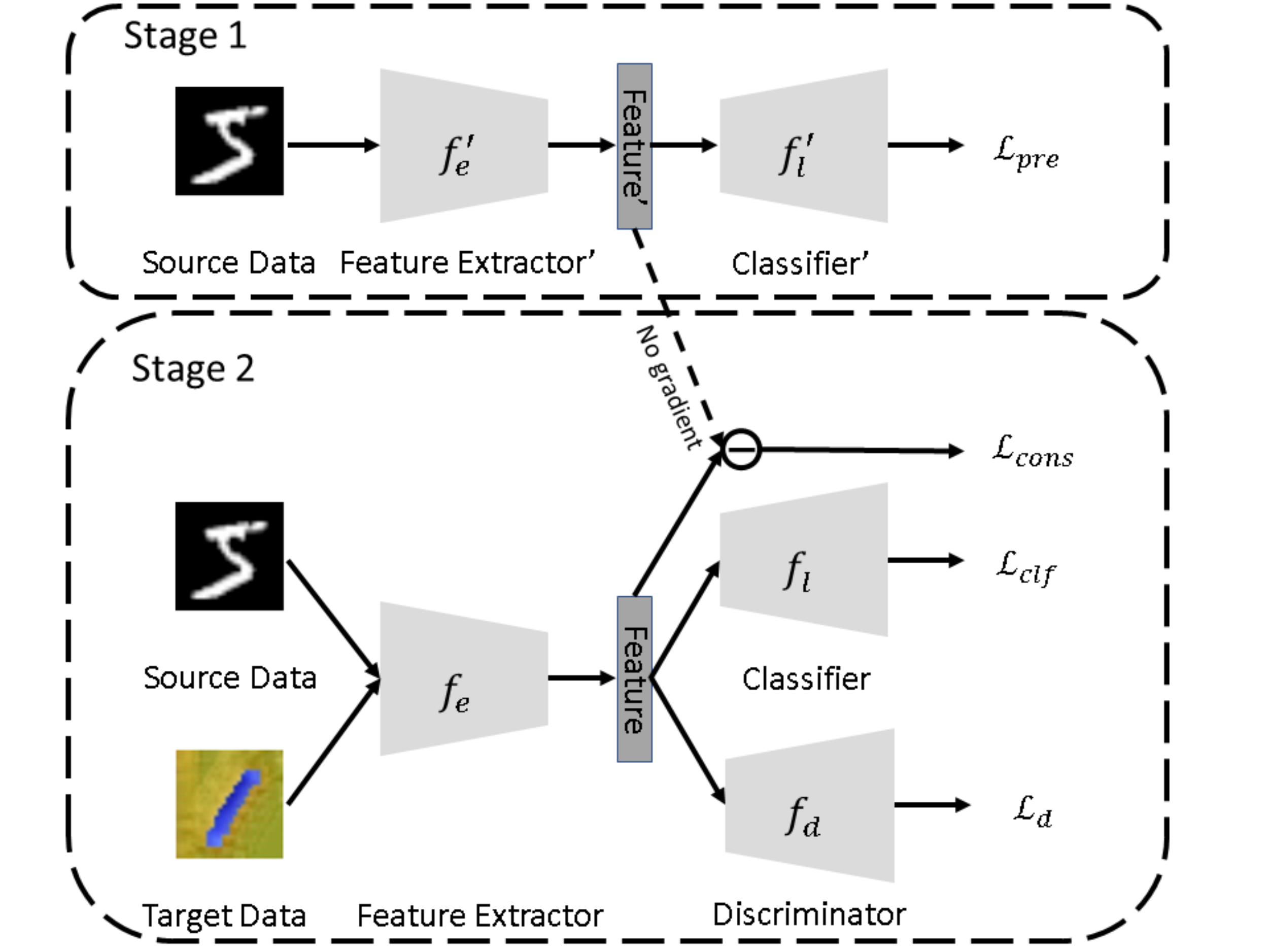}}
	
	\caption{The proposed two-stage unilateral alignment method. (Stage 1) We extract source features using a separate network. (Stage 2) We unilaterally align the distributions by adding the consistency loss. The loss is computed using current calculated source features and its corresponding pre-trained features.}
	\label{stage}
\end{figure}

We visualize the proposed method in Figure~\ref{stage}.

\subsection{Stage 1: Extract Inter-class Relationship via Source-only Learning} 
Since the missing classes are unavailable for the target domain, it is only possible to learn inter-class relationships from source data. We thus propose to pre-train a separate neural network to extract this relationship. We assume that we have the same backbone architecture as the main network used in the second stage: a feature extractor $f'_e$ parameterized by $\theta'_e$, and a classifier $f'_l$ parameterized by $\theta'_l$.  We, thus, apply a standard supervised training procedure to learn the relationship. Formally, we train this separate network by using the following loss function: 
$$\mathcal{L}_{pre} = \mathcal{L}_{clf_{\mathcal{D}_s}} (\theta'_e, \theta'_l),$$
where $\mathcal{L}_{clf}$ is the softmax cross entropy loss~\cite{bishop2006pattern, goodfellow2016deep} function, which is widely used in supervised classification problems. The network is trained only source data 
$\mathcal{D}_s$. This stage-1 network is frozen after this stage of training. 

After a successful training on the network in stage 1, we can extract the pre-trained source features $f'_e(x_{si})$ for each source training sample $x_{si}$. These features contain meaningful inter-class relationships because a simple classifier is able to make strong predictions for source data. These features are then used as a reference for the next training step. 

\subsection{Stage 2: Unilateral Adversarial Domain Adaptation} 
\subsubsection{Off-the-shelf Adversarial Domain Alignment}
Motivated by the successful applications of DANN in computer vision~\cite{ganin2014unsupervised, ganin2016domain} and its later adoption in industrial application~\cite{wang2019domain}. We propose to make use of this off-the-shelf adaption technique in our adaptation task with missing classes. 

As shown in Figure~\ref{ugrl}, our main architecture has three components: the feature extractor $f_e$, the classifier $f_l$, and an additional discriminator $f_d$.   The alignment is achieved by introducing the discriminator to distinguish between healthy features coming from the source and those from the target. Meanwhile, the feature extractor is encouraged to fool the discriminator so that the features are unbiased towards their origin. Formally, this is equivalent to the following mini-max problem~\cite{ganin2014unsupervised}:

$$\mathcal{L}(\theta_e,\theta_l,\theta_d) = \mathcal{L}_{clf_{\mathcal{D}_s}} (\theta_e, \theta_l) - \lambda_d \mathcal{L}_{d_{\mathcal{D}_s, \mathcal{D}_t}}(\theta_e, \theta_d),$$
$$(\hat\theta_e, \hat\theta_l) = \arg\min_{\theta_e, \theta_l} \mathcal{L}(\theta_e, \theta_l, \hat\theta_d),$$
$$\hat\theta_d = \arg\max_{\theta_d} \mathcal{L}(\hat\theta_e, \hat\theta_l, \theta_d),$$
where $\mathcal{D}_s$ is the source data and $\mathcal{D}_t$ is the target data,  $\mathcal{L}_{clf}$ is again the softmax cross entropy loss function. $\mathcal{L}_{d}$ is the cross entropy loss for the domain classification subtask. The objective function is similar to that of Generative Adversarial Networks (GAN)~\cite{goodfellow2014generative}. It includes two parts: a classification loss for supervised learning and a domain adversarial loss for alignment.

We minimize the classification loss, w.r.t. parameters of the feature extractor and classifier. In addition to this supervised loss, we maximize the adversarial alignment loss w.r.t. parameters of the feature extractor in order to achieve domain-invariant features. We further minimize the adversarial alignment loss w.r.t. the discriminator, and thus, train the discriminator to provide a precise prediction of the origin of features. 

It has been shown in~\cite{ganin2014unsupervised} that by reversing feature extractor's gradients before sending them to the domain classifier, we can reformulate the problem and alleviate the $\mathcal{H}$-Divergence between source and target distributions~\cite{ben2010theory, cortes2011domain}. We adopted this optimization technique for our domain adaptaion problem with missing classes. Using the gradient reverse layer (GRL) from DANN~\cite{ganin2014unsupervised, ganin2016domain}, the above loss function can be rewritten as:
$$\mathcal{L}_{align}  = \mathcal{L}_{clf} + \mathcal{L}_{d}.$$
Thus, it is able to be trained as an end-to-end learning problem. 
\begin{algorithm}[t]
\SetAlgoLined
\SetKwInOut{Input}{Input}
\SetKwInOut{Output}{Output}
\caption{Training procedure for the proposed unilateral alignment method.}
\label{algo:alg}
\Input{Source and target training samples.}
\begin{enumerate}

  \item Stage 1: Train a source feature extractor $f_e'$ and classifier $f_l'$ using only source data, then freeze this pre-trained network. Extract reference source features $f_e'(x_s)$ from the pre-trained network.  
  \item Stage 2: Unilateral Adversarial Domain Adaptation
  \begin{enumerate}
  \item Initialize a new feature extractor $f_e$, a new classifier$f_l$, and a domain discriminator$f_d$. 
  \item Calculate the classification loss $\mathcal{L}_{clf}$ using source data.
  \item Calculate the consistency loss $\mathcal{L}_{cons}$ using source data and pre-trained features. 
  \item Reverse gradients for healthy features from both source and target, then calculate the adversarial alignment loss $\mathcal{L}_{d}$
  \item Calculate the overall loss and gradients, update $\theta_{e}, \theta_{d}, \theta_{l}$ accordingly. 
  \item Go back to b) for next iteration training until convergence.
  \end{enumerate}
  \item Evaluate the learnt model on the target test set with all classes. 
\end{enumerate}

\end{algorithm}
\subsubsection{Unilateralism as an Additional Loss Term} 
The DANN method is directly aligning the complete source data with a target data that has missing classes. Thus a significant misalignment is expected. In order to avoid the potential negative effect of the above alignment and preserve the inter-class relationships while applying domain adaptation techniques, we propose to unilaterally align the target distribution to the corresponding part of the source distribution, instead of aligning both to a shared new space. We consider the pre-trained source features as a good representation of all classes, since class separability could be achieved. To transfer this good representation, in stage 2, an additional constraints is applied to force the aligned source features to be as close as possible to the pre-trained one. If the alignment is successful, then the target features should also be aligned with the pre-trained source features.

In order to preserve the inter-class relationships while applying the partial domain-adversarial alignment, we make use of the pre-trained source features $f'_e(x_{s})$, and force the aligned source features to be close to the pre-trained ones:
$$\mathcal{L}_{cons} =\frac{1}{K} \sum_{j=1}^{K} ||f(x_{s})_j - f'(x_{s})_j||_1, $$
where K is the number of features in the feature space. We add this additional constraint to the loss function described in the previous paragraph. The overall loss function becomes thereby:
$$\mathcal{L}  = \mathcal{L}_{clf} + \mathcal{L}_{d} + \lambda_{cons}\mathcal{L}_{cons} .$$

This additional loss is inspired by the consistency loss introduced in~\cite{chen2018domain} where a similar distance is used to improve the cross-domain robustness of the bounding box predictor for object detection tasks. However, it is used for a different purpose here since we are trying to encourage the alignment in one direction and preserve the inter-class relationships.  We tested the additional loss using both $l_1$ and $l_2$ loss and found no significant difference between them.

\subsection{Summary}
To summarize, in addition to aligning the source and target distributions via DANN, we propose to impose an additional constraint to make the alignment unilateral towards the pre-trained source features. The main objective of the unilateral alignment is to preserve the inter-class relationships learned from the source data, where knowledge on all classes are available.

\section{Experiments on MNIST $\rightarrow$ MNIST-M Task}
\begin{figure}
    	\centering
	{\includegraphics[width=0.5\columnwidth]{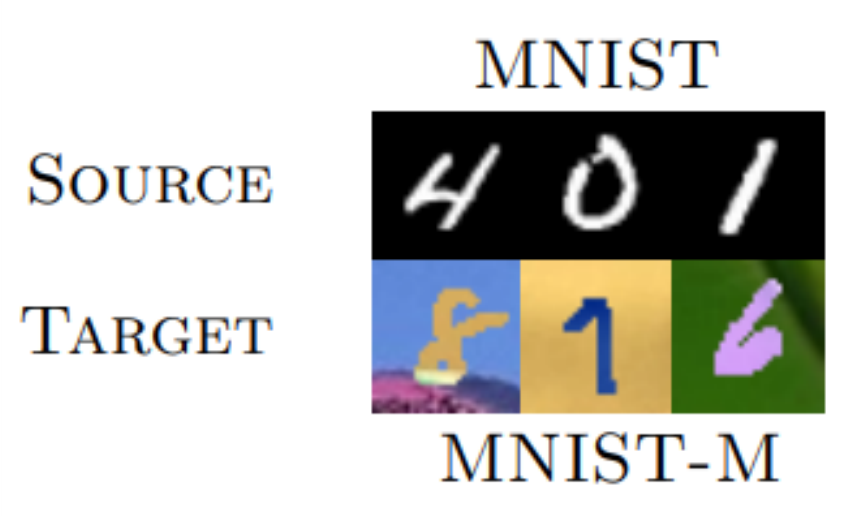}}
	
	\caption{Training samples from the MNIST$\rightarrow$ MNIST-M task. Figure taken from~\cite{ganin2016domain}. }
	\label{mnistm}
\end{figure}
Following the experiment setup used by~\cite{ganin2016domain}, we first evaluate our model using the MNIST~\cite{lecun1998gradient}$\rightarrow$ MNIST-M~\cite{ganin2016domain} task~\footnote{
		Our code for this experiment is available at \url{https://github.com/diagnosisda/dxda}}. The MNIST-M dataset blends digits from MNIST over patches randomly extracted from color photos from BSDS500~\cite{arbelaez2010contour}. Samples from both datasets are shown in Figure~\ref{mnistm}. It is a classification task with 10 digits $0\rightarrow 9$ as 10 classes. The publicly-available implementation$\footnote{https://github.com/pumpikano/tf-dann}$ of DANN  is used for our MNIST experiments. 

We evaluate the proposed unilateral alignment model against a source-only baseline and the DANN alignment without unilateral constraint. We conduct experiments  on different level of missing classes as shown in Table~\ref{mmnisttable}. For example, in the $8/10$ setup, the unlabeled target training data contains  samples from label $[0, 1]$. In the $2/10$ setup, the unlabeled target training data contains samples from label $[0,1,2,3,4,5,6,7]$. The source-only baseline is trained for 10 Epochs. Both DANN and DANN+Unilateral are trained for 50 Epochs. We adopt the training parameters used by~\cite{ganin2016domain}. We use a batch size of 64 and learning rate of 0.01. 

Results on this task are shown in Table~\ref{mmnisttable}. All reported results are based on the average accuracy of five different runs. As shown in the third column in the table, without the unilateral constraint, the benefit of the alignment is significantly decreasing when there are more missing classes.  This is because the misalignment effect is larger when more  classes are missing. By simply adopting our additional unilateral loss, the inter-class relationship learned from the source domain is strengthened, and the adaptation is significantly improved. When there are 9 out of 10 classes missing in the target training set, the unilateral model performance is less than $4\%$ worse than the full alignment, comparing to the $\approx 50\%$ degradation without our additional loss. This demonstrates the effectiveness of our proposed method.

Note that in the last row, when there is no missing classes, the additional unilateral loss is limiting the adaptation ability of DANN, thus slightly hurts the model performance. 

\begin{table}[]
\caption{Experiment results on the MNIST$\rightarrow$ MNIST-M task.}
\begin{tabular}{|c|c|c|c|}
\hline
\# of missing class  & Baseline & DANN~\cite{ganin2016domain}    & DANN+Unilateral (Ours) \\ \hline
9 / 10                         & 57.49\%   & 27.03\% & \textbf{72.99}\%     \\ \hline
8 / 10                         & 57.49\%   & 54.72\% & \textbf{74.06}\%     \\ \hline
7 / 10                         & 57.49\%   & 72.76\% & \textbf{74.72}\%     \\ \hline
6 / 10                         & 57.49\%   & 72.97\% & \textbf{74.40}\%     \\ \hline
5 / 10                         & 57.49\%   & 73.91\%  & \textbf{74.99}\%     \\ \hline
2 / 10                         & 57.49\%   & 74.76\% & \textbf{75.03}\%     \\ \hline
0 / 10                        & 57.49\%   & \textbf{76.95}\%    & 76.74\%     \\ \hline
\end{tabular}
\label{mmnisttable}
\end{table}


\section{Experiments on Fault Diagnosis Problems}
In the following section, we demonstrate the benefits of the proposed approach on a different fault diagnosis task. Fault diagnosis is a classification task where our method can be directly applied on. Usually the label space consists of healthy and fault conditions. In this section, we consider the following variant of fault diagnosis problem: 

Unsupervised domain adaptation when $80\%$ of classes are missing in the target training set.

\begin{table}[b]
\caption{Class Definition for the CWRU Dataset~\cite{smith2015rolling}. Table taken from~\cite{wang2019domain}.}

    \begin{center}
\begin{threeparttable}[b]
        \begin{tabular}{|c|c|c|c|c|c|c|c|c|c|c|}
        \hline
        \textbf{Fault }&\multicolumn{10}{c|}{\textbf{Class Label}} \\
        \cline{2-11} 
        \textbf{} & 0& 1& 2& 3& 4& 5& 6& 7& 8& 9 \\
        \hline
        Loc&NA$^1$& IF& IF&  IF&  BF&BF&BF&OF&OF&OF  \\
        \hline
        Size&0& 7& 14&  21&  7&14&21&7&14&21  \\
        \hline
        \end{tabular}

 \begin{tablenotes}
     \item[1] Fault location not applicable because class 0 is the healthy state.
   \end{tablenotes}
  \end{threeparttable}

    \end{center}
\label{tab1}
\end{table}

\subsection{Dataset}
We conduct our experiments on the fault diagnosis dataset: the Case Western Reserve University (CWRU)~\cite{smith2015rolling}.

\subsubsection{CWRU}
The publicly-available bearing dataset from CWRU is used. It is a widely used dataset in the field of domain adaptation for fault diagnosis~\cite{li2018cross, lu2016deep, zhang2018adversarial, wang2019domain}.

We follow the setup used by~\cite{li2018cross} whenever possible. Thus drive-end accelerometer data are used as our input. The list of labels concerned in this paper is shown in Table~\ref{tab1}, namely three different fault types, along with one healthy state are considered. IF stands for Inner race fault, BF stands for Bearing fault, and OF stands for outer race fault. Each fault type contains three sub-types, with fault diameters of 7, 14, 21 mils. Sampling rate of 12 kHz is adopted. Whenever data are unavailable at 12 kHz, we down-sample them to ensure a consistent sampling rate of 12 kHz in all experiments. 

There are four different loads $\{0, 1, 2, 3\}$ in the CWRU dataset. Domain adaptation is applied across these four different loads. For example, Task $0\xrightarrow{}1$ means working load 0 is the source domain with labeled training samples, and working load 1 is the target domain we want to improve model performance on.

\subsection{Pre-processing}
\begin{figure}
    	\centering
	{\includegraphics[width=\columnwidth]{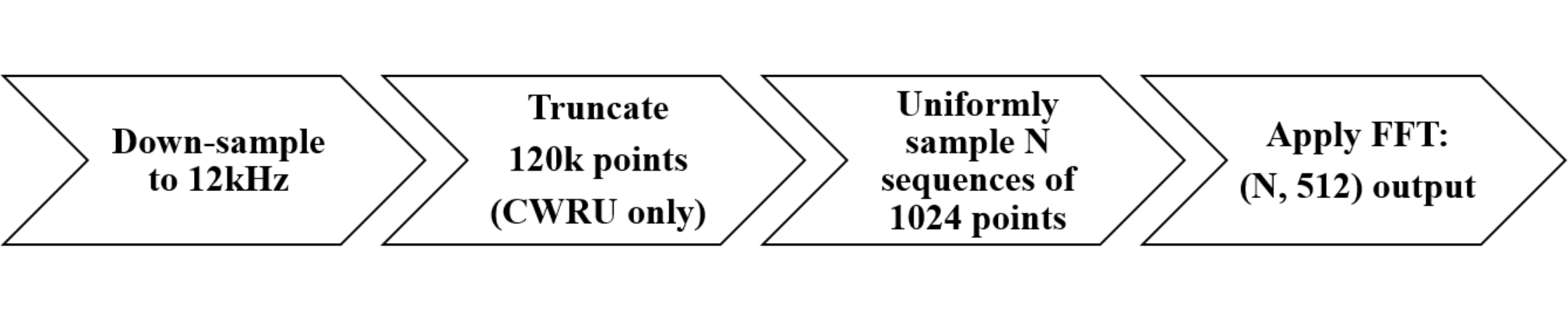}}
	
	\caption{Preprocessing steps taken from~\cite{li2018cross, wang2019domain}.}
	\label{flowchart}
\end{figure}

For the CWRU dataset, we follow the same pre-processing steps as in~\cite{li2018cross, wang2019domain}. As shown in Figure~\ref{flowchart}, first, we downsample and truncate each raw recording. Second, we divide each it into 200 sequences of 1024 points. Finally, using the Fast Fourier Transform~\cite{cooley1965algorithm}, each sequence is converted into a vector of 512 Fourier coefficients.

\subsection{Model Implementation}
\begin{figure}
    	\centering
	{\includegraphics[width=\columnwidth]{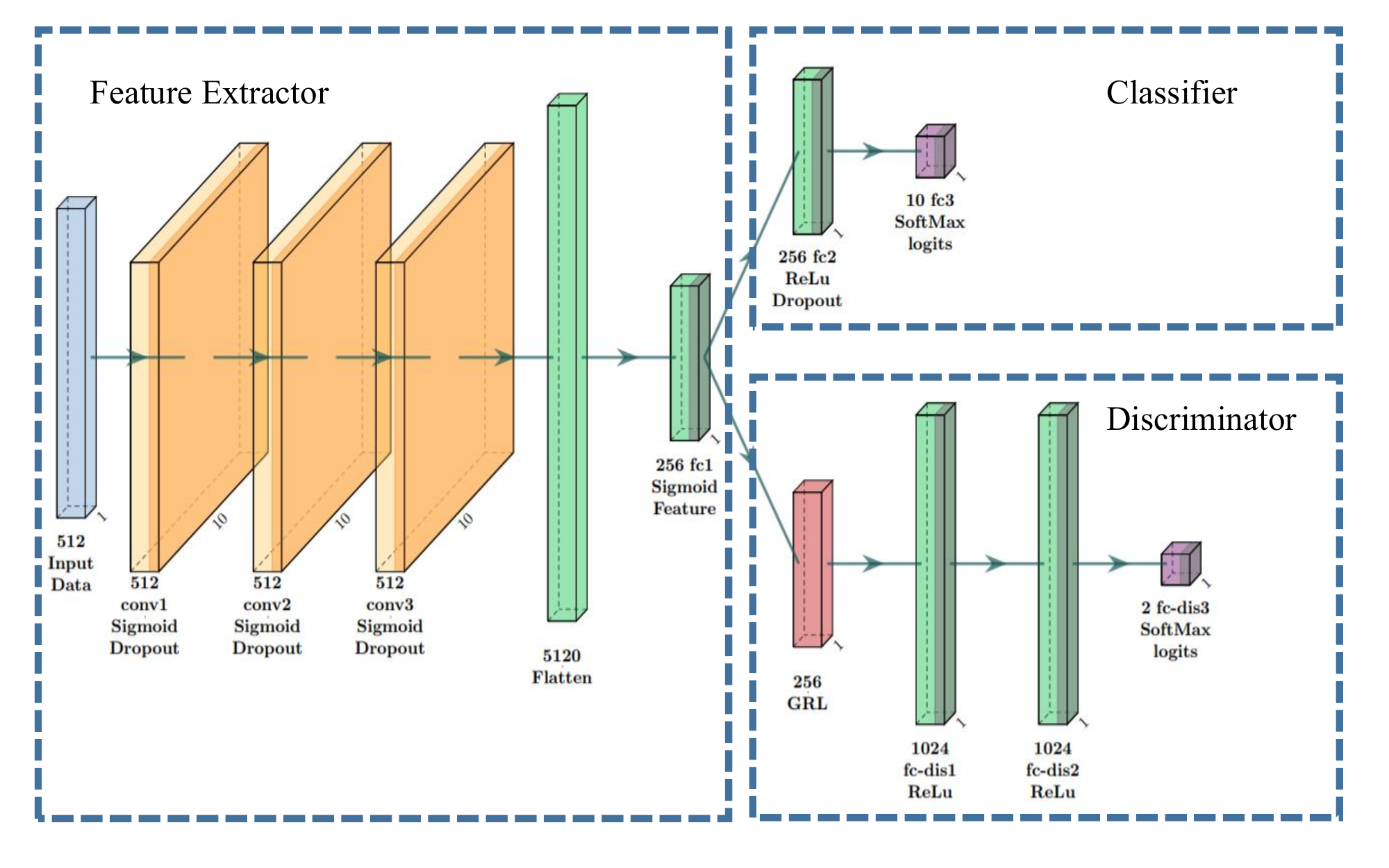}}
	
	\caption{Backbone network used in all our fault diagnosis experiments. The feature extractor and classifier are taken from~\cite{li2018cross, wang2019domain}.}
	\label{ugrl}
\end{figure}

We visualize the details of the backbone model and our discriminator in Figure~\ref{ugrl}. We use the same architecture following~\cite{li2018cross, wang2019domain} to enable a fair comparison.

The backbone architecture~\cite{li2018cross} contains two components: a feature extractor, and a classifier. Each convolutional layer has a filter length of 3, and a hidden size of 10. Dropout layers are added accordingly afterwards with a dropout rate of 0.5. The signal is then flattened and transformed to features of size 256 by a fully-connected layer. The classifier is a single hidden layer network of size 256, using the softmax cross-entropy loss.

We train this backbone network as our baseline model using source load data only. To implement our models, a discriminator is needed additionally which comprises two fully connected hidden layers followed by the softmax cross-entropy loss.

The architectures of the backbone network and of the additional discriminator are illustrated in Fig~\ref{ugrl}. The CWRU models are trained for 2000 Epochs with Sigmoid activation function. We report 5-run average accuracy and standard deviation. 

\subsection{Experiment Results}
\subsubsection{Unsupervised Domain Adaptation Experiment on CWRU Dataset with 80\% Missing Classes}

In this experiment, the target training dataset is composed of a subset of the classes. We would like to again emphasize that none of the target class labels are used for training.  Formally,  we consider the following set of unlabeled training data from the target machine $$\mathcal{D}_{t}=\{(x_{t}^1, y_{t}^1), ...,(x_{t}^m, y_{t}^m)\},  y_{t}^i  \in Y_{sub}=\{0, ..., k-1\}.$$ for the missing classes experiments. We consider $k=2$ for our demonstration.

Note that in this setup we choose arbitrarily the first 2 classes of the dataset ($Y_{sub}=\{0, 1\}$) to demonstrate the effectiveness of our unilateral alignment approach, but one may expect similar behavior when the subset of fault types is chosen differently.

\begin{table}[]
\centering
\caption{Unsupervised Domain Adaptation Experiment Result on CWRU Dataset with 80$\%$ Missing Classes on CWRU Dataset}

   \begin{threeparttable}[b]
\begin{tabular}{|c|c|c|c|}
\hline
Setup    & Baseline${^1}$      & DANN~\cite{ganin2016domain}${^{1,2}}$          & DANN+Unilateral${^{1,2}}$          \\ \hline
Task 0-1 & 93.49 $\pm$ 1.75  & 92.40 $\pm$  1.84  & \textbf{96.36 $\pm$  0.81} \\ \hline
Task 0-2 & 93.65 $\pm$ 0.96  & 92.86 $\pm$  1.53  & \textbf{97.38 $\pm$  1.66} \\ \hline
Task 0-3 & 91.02 $\pm$ 1.92  & 92.84 $\pm$  1.66  & \textbf{95.78 $\pm$  1.90} \\ \hline
Task 1-0 & \textbf{97.93 $\pm$ 0.24}   & 96.92 $\pm$  0.74  & 97.49 $\pm$  0.52\\ \hline
Task 1-2 & \textbf{100.00 $\pm$ 0.00} & 99.96 $\pm$  0.06 & 99.94 $\pm$  0.04 \\ \hline
Task 1-3 & 98.26 $\pm$ 1.63  & 99.50 $\pm$  0.09  & \textbf{99.58 $\pm$  0.15} \\ \hline
Task 2-0 & 91.63 $\pm$ 1.82  & 93.49 $\pm$  0.79  & \textbf{93.77 $\pm$  2.09} \\ \hline
Task 2-1 & 97.09 $\pm$ 0.92  & 97.56 $\pm$  0.19  & \textbf{97.60 $\pm$  0.36} \\ \hline
Task 2-3 & 99.78 $\pm$ 0.17  & \textbf{99.90 $\pm$  0.09}  & 99.86 $\pm$  0.08 \\ \hline
Task 3-0 & 87.96 $\pm$ 0.18  & 88.41 $\pm$  0.22  & \textbf{88.42 $\pm$  0.72} \\ \hline
Task 3-1 & 89.42 $\pm$ 0.96  & 90.53 $\pm$  1.14  & \textbf{93.45 $\pm$  1.68} \\ \hline
Task 3-2 & 99.65 $\pm$ 0.17  & 99.14 $\pm$  0.90  & \textbf{99.83 $\pm$  0.06} \\ \hline
Average     & 94.99         & 95.29           & \textbf{96.62}          \\ \hline
\end{tabular}
\label{tab:50}
   \begin{tablenotes}
     \item[1] Reported numbers are based on average and standard deviation over five runs. 
     \item[2] In training phase, labeled source samples from all classes and unlabeled target samples from the first 20\% classes are provided. In test phase, target samples from all classes, including those that are unseen in target during training are evaluated.  
   \end{tablenotes}
  \end{threeparttable}
\end{table}

The results under this new setup are shown in Table~\ref{tab:50}, for the source only baseline, DANN and DANN with our additional unilateral constraint.
Compared to the baseline, DANN alone does not provide a significant improvement. This is likely due to the negative effect of trying to align the source features of all classes  with target features containing only 20\% of the classes. This could result in a distortion of the inter-class relationships. 

The proposed unilateral alignment method, tackles this negative effect and strengthens the alignment. By adding the additional consistency loss, it provides an additional 1.33\% absolute accuracy improvement over the naive implementation of DANN.

\subsubsection{Discussion}
The performed fault diagnosis experiments on the case study demonstrate that the unilateral alignment is able to improve model performance of domain adaptation problems, when there are missing classes in the target training set. 

There are few cases where we observe that unilateral alignment may slightly hurt the performance. In such cases, results show that the alignment was actually not required: the baseline is already providing a very high accuracy. Our results demonstrate that in such cases, the drop in accuracy is very small or even insignificant while in many other cases, the gain in accuracy is  significant. Overall, the results are significantly improved by the proposed methodology.

\section{Conclusions}
In this paper, we demonstrated that when there are missing classes in the target training dataset, directly applying adversarial domain adaptation techniques results in performance decrease.  To overcome this problem, we proposed the unilateral alignment, a simple yet effective training strategy that leverages the inter-class relationships of the source domain.  We showed in the MNIST experiment that by adding the additional consistency loss that enforces the unilateral alignment the model is able to be robust against missing classes. The additional experiments on fault diagnosis tasks show a promising potential of the proposed  domain adaptation method for industrial applications where the problem of missing classes imposes a significant limitation on the applied approaches. Exploring the performance of the proposed models in case  of corrupted samples is one of the future directions. 

\section*{Acknowledgement}
The authors thank Prof. Wen Li for fruitful discussions.

\bibliographystyle{Bibliography/IEEEtranTIE}
\bibliography{Bibliography/IEEEabrv,Bibliography/bib}\ 

\end{document}